\documentclass{aa}
\usepackage[varg]{txfonts}
\usepackage{graphicx}
\usepackage{natbib}
\bibpunct{(}{)}{;}{a}{}{,} 
\usepackage{color}

\begin{document}

\title{Characterisation of chaos in meteoroid streams.}
\subtitle{Application to the Geminids.}
\author{Ariane Courtot \thanks{ariane.courtot@obspm.fr} \and Jérémie Vaubaillon \and Marc Fouchard}
\institute{IMCCE, Observatoire de Paris, PSL Research University, CNRS, Sorbonne Universit\'{e}, UPMC Univ Paris 6, Univ Lille, France}
\date{Received 20 October 2022 / Accepted date 28 March 2023}

\abstract
{Dynamically linking a meteor shower with its parent body can be challenging. This is in part due to the limits of the tools available today (such as D-criteria) but is also due to the complex dynamics of meteoroid streams.}
{We choose a method to study chaos in meteoroid streams and apply it to the Geminid meteoroid stream.}
{We decided to draw chaos maps. Amongst the chaos indicators we
studied, we show that the orthogonal fast Lyapunov indicator is particularly well suited to our problem. The maps are drawn for three  bin sizes, ranging from $10^{-1}$ to $10^{-4}$ m.}
{We show the influence of mean-motion resonances with the Earth and with Venus, which tend to trap the largest particles. The chaos maps present three distinct regimes in eccentricity, reflecting close encounters with the planets. We also study the effect of non-gravitational forces. We determine a first approximation of the particle size $r_{lim}$ needed to counterbalance the resonances with the diffusion due to the non-gravitational forces. We find that, for the Geminids, $r_{lim}$ lies in the range $[3;8]\times 10^{-4}$ m. However, $r_{lim}$ depends on the orbital phase space.}
{}

\keywords{Gravitation -- Chaos -- Methods: numerical -- Celestial mechanics -- Meteorites, meteors, meteoroids} 

\maketitle

\renewcommand{\thefootnote}{\fnsymbol{footnote}}
 
\section{Introduction}
The Meteor Data Center of the International Astronomical Union (IAU) currently lists 921 meteor showers\footnote[2]{https://www.ta3.sk/IAUC22DB/MDC2007/ visited in September 2022}, most of which are in the working list and are awaiting confirmation or more data. According to this number, on average, $2.6$ near-Earth objects per day were active enough in the past $10^3-10^4$ years to produce such showers. If this is confirmed, it would greatly impact our current understanding of the Solar System. This prompted us to examine how IAU meteor showers are determined.

Most methods used to find new meteor showers involve computing the radiant and an orbit dissimilarity criterion (D-criterion). D-criteria quantify the proximity of the orbits of two objects. Many D-criteria have been developed \citep[e.g.][]{Valsecchi_al_1999, Jenniskens_2008,Rudaska_al_2015} but the $D_{SH}$ from \citet{Southworth_Hawkins_1963} is still largely used today. However, this criterion has long been criticised for its mathematical, physical, and statistical shortcomings \citep[see e.g.][]{Drummond_1981, Valsecchi_al_1999}. 

\citet{Rudawska_Jopek_2010}  compared two criteria (the $D_{SH}$ and the criterion described by \citealp{Jopek_al_2008}), providing some preliminary indications as to the validity of the two. However, to our knowledge, no such study has been conducted for the entire set of D-criteria: it is not known which should be avoided and which are most suited to specific cases.

While D-criteria are not sufficient to identify a meteor shower with absolute certainty, they can help us to identify candidates. We call those candidates meteor groups: {a meteor group is a set of meteors sharing a radiant and showing similar orbits}. In contrast, meteor showers are defined, according to the IAU, as {a set of meteors coming from a single parent body, through a meteoroid stream}\footnote[3]{"Definitions of Terms in Meteor Astronomy" from https://www.iau.org/science/scientific\_bodies/commissions/F1/ visited in October 2022}. To prove that a given meteor group is in fact a meteor shower, a statistical or dynamical analysis is sometimes performed \citep[see e.g.][]{Guennoun_al_2019} but these are not always conclusive.

Most dynamical analyses model meteoroids from a suspected parent body and follow these particles through time until they meet with the Earth \citep[see e.g.][]{Egal_al_2021}. However, dynamical chaos has not been studied extensively, perhaps because of the specificity of meteoroid dynamics (mainly the non-gravitational forces that make for a non-conservative problem).

Chaos could explain the difficulty in understanding the dynamical evolution of the meteoroids and provide insights into the formation of the meteoroid streams. The study of chaos is often carried out using chaos maps. These maps can be found as far back as 1990 \citep{Markus_1990}, and they have become a standard means to describe chaotic and stable regions of a phase space as a function of initial orbital elements. They are generally drawn using a chaos indicator derived from the theory on Lyapunov characteristic exponents \citep{Benettin_al_1980}, and usually for a specific type of body (e.g. moons or asteroids). Here, we use this technique to explore meteoroids, and this contributes to a new field of study.

Combined with other tools (radiant, D-criteria, statistical and dynamical analysis), chaos maps could help us to prove whether or not a meteor group is a meteor shower. For example, if a meteor group is shown to come from a part of the map that never intersects the Earth, it cannot be a meteor shower, as no parent body would explain that dynamic. Another example would be a meteor group with a very small D-criterion in a very chaotic region. If this hypothetical meteor group was a meteor shower, its meteoroids would be scattered quickly because of the chaos. Such a meteor shower would not survive for a long time, and so if observations of this meteor group date back sufficiently far in time, it is very unlikely to be a meteor shower. 

Our aim is not to develop a new method to study chaos, but to use well-known tools in a new field of study. We can expect most meteoroids to be chaotic, as they are subjected to many close encounters and are under the influence of non-gravitational forces; our aim is to precisely quantify this chaos and to investigate what drives the dynamics of meteoroids.

In Sect. \ref{sec:meth}, we outline the method used to draw meteoroid stream chaos maps, and we explain our choice of chaos indicator. In Sect. \ref{sec:results}, we present an application of this method to the Geminid meteoroid stream. More precisely, the maps drawn for the Geminids show which resonances constrain the evolution of the meteoroids. We also investigate the impact of non-gravitational forces and study the role of eccentricity in the definition of the Geminids.

\section{Method}\label{sec:meth}

\subsection{Chaos indicator}\label{subsec:indicator}

Chaos maps are drawn using an indicator that measures the chaoticity of a given orbit. Various indicators and methods \citep[such as the analysis in frequency from][]{Laskar_1990} are available to study chaos.
Here, the indicator has to be suitable for meteoroid analyses: meteoroid evolution is characterised by the effects of non-gravitational forces. These forces greatly reduce the timescale of meteoroid evolution, meaning that these objects survive for a few thousand years at most \citep{Liou_Zook_1997}; they also make the problem non-conservative.

The most common chaos indicators are based on the divergence of two initially nearby orbits: a chaotic behaviour is characterised by an exponential divergence, in contrast to the linear divergence of a stable behaviour. The Lyapunov characteristic exponents are based on this idea, but they use tangent vector and variational equations instead of nearby orbits, as described in Subsect. \ref{subsubsec:formula}. These Lyapunov characteristic exponents are the basis of the indicators we study here.

These chaos indicators are either relative or absolute. The former are generally used to draw a map over a wide area of the phase space, while the latter are usually used to study a specific object. Most chaos indicators could potentially be suitable for our problem. For example, we could have considered indicators such as the one proposed by \citet{Barrio_2005}, but we restricted ourselves to only a couple of Lyapunov-based indicators, as a complete study is beyond the scope of this paper.

\subsubsection{Choice of indicator}\label{subsubsec:choice}

We compared the fast Lyapunov indicator (FLI) described in \citet{Froeschle_al_1997}, the modified FLI (mFLI) from \citet{Guzzo_Lega_2015}, the mean exponential growth factor of nearby orbits (MEGNO) and mean MEGNO (mMEGNO) from \citet{Cincotta_al_2003}, and the orthogonal fast Lyapunov indicator (OFLI) from \citet{Fouchard_al_2002}. Below, we explain why we chose this last indicator. 

All indicators have strong arguments in favour of their study and they have all been used successfully. However, none of them were initially developed for short timescales (order of $10^3$ years) or for objects under the influence of non-gravitational forces. It is therefore necessary to verify whether or not they are suitable to our problem.

The mFLI describes close encounters, which play an important role in the dynamics of meteoroids. However, this indicator is designed to study a specific encounter, and not the general effect of these encounters on a relatively large region of the phase space. A discussion with the authors led us to realise that this indicator was not adapted to our problem.

The FLI, mFLI, and OFLI are relative: they measure the chaoticity of an orbit, but only relative to others. The lower their value, the more stable the orbit studied. The MEGNO is the only absolute indicator presented here, with a value of two being the threshold between chaos ($>2$) and stability ( $ \leq 2$). However, its oscillations make it ill-suited to a map, as its value at a time $t$ might not represent its general behaviour. Nevertheless, mMEGNO corrects for this problem, and seems to be generally preferred over MEGNO for drawing maps.
Comparing FLI and OFLI, the latter filters out an artificial effect: the growth of the FLI due to differential rotation. 

\begin{figure}
    \includegraphics[scale = 0.5]{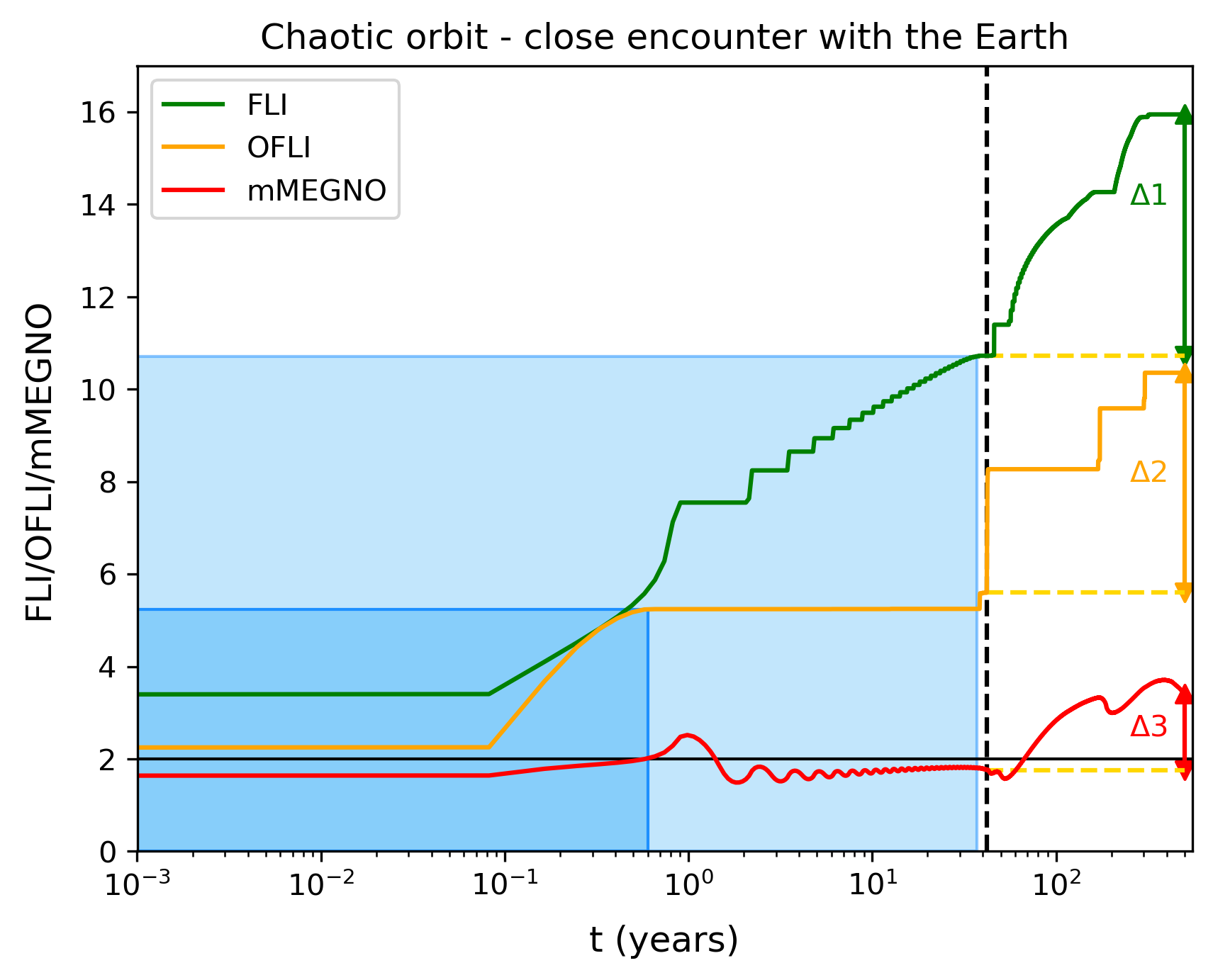}
    \includegraphics[scale = 0.5]{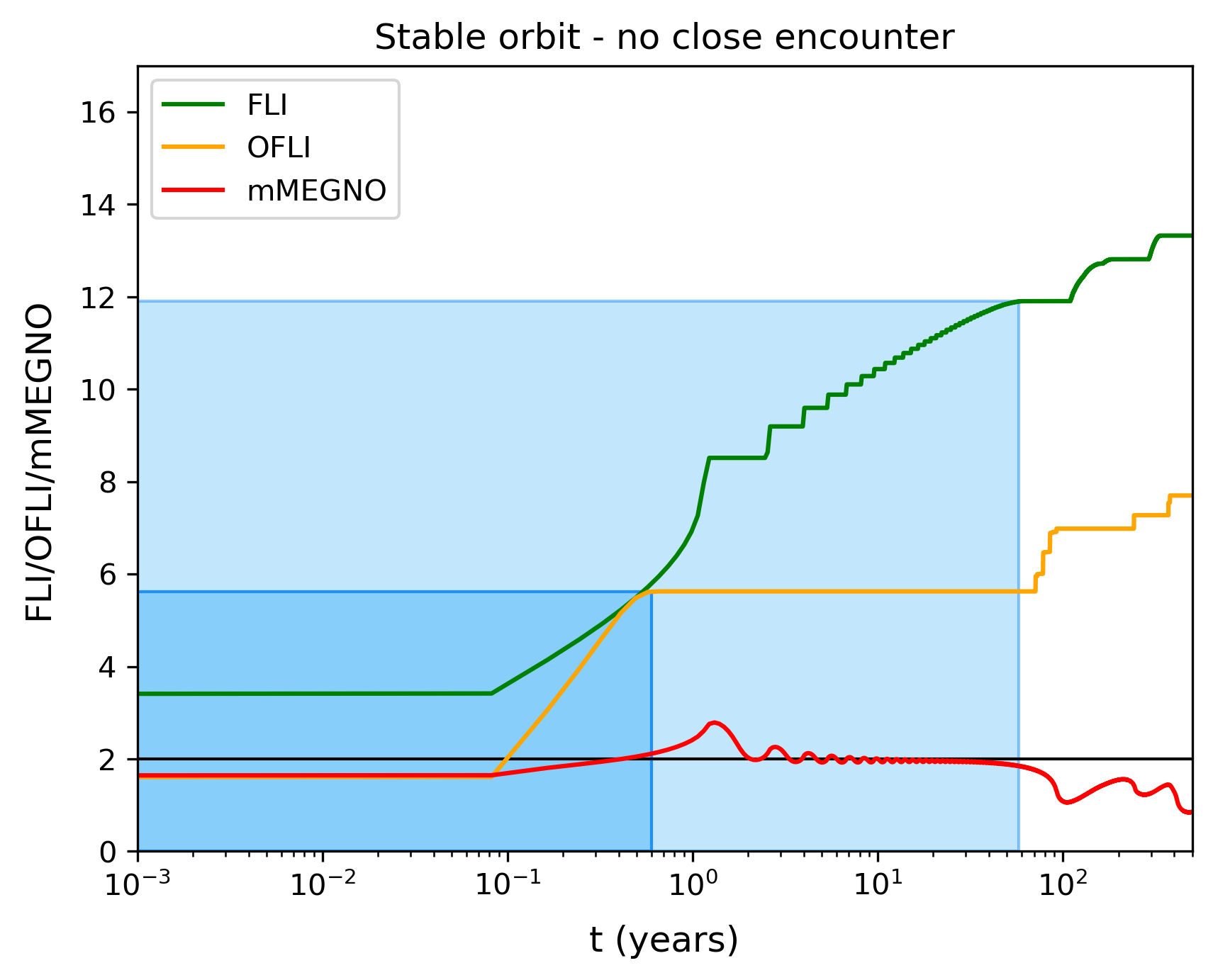}
    \caption{Comparison between the evolution of FLI, OFLI, and mMEGNO for two particles, a chaotic particle (mMEGNO > 2) with a close encounter with the Earth, and a stable particle (mMEGNO < 2) without close encounters. The rectangles show the initialisation phase for the FLI (larger, very light blue) and for the OFLI (smaller, light blue). The vertical black dashed line marks the close encounter of the chaotic particle with the Earth. The horizontal yellow dashed lines mark the values of the indicators immediately before the encounter. These values are compared with the final value of each indicator thanks to the arrows. The precise values of $\Delta_1$, $\Delta_2$, and $\Delta_3$ are given in the text, with the overall interpretation.}
    \label{fig:comp_indic}
\end{figure}

We compared FLI, OFLI, and mMEGNO, examining the evolution of the indicators during the integration of 12 particles from the Geminid meteoroid stream. This integration lasts 500 years. We only took into account the gravitational forces from all the planets. Figure \ref{fig:comp_indic} shows the comparison of the indicators for two particles: one that experienced a close encounter and therefore became chaotic, and another that remained stable and did not encounter a planet. We would like to point out that this second particle is a rare case, and is only presented here for comparison, as the large majority of our particles are unstable.

The blue rectangles in Fig. 1 represent the initialisation phase of the FLI (light blue) and of the OFLI (darker blue). The initialisation phase ends when the value of the indicator first levels off. The OFLI has a very short initialisation phase: it reaches its first plateau in 0.6 years, whereas that of the FLI takes between 37 and 58 years depending on the particle (between 7.4\% and 11.6\% of the total integration time). The definition of this initialisation phase is less clear for the mMEGNO.

For the first particle, a black dashed line marks a close encounter with the Earth (the particle's distance to Earth was smaller than its Hill radius). We wanted to test the impact of a close encounter because meteoroids are characterised by their numerous encounters with planets. This close encounter happened just after the initialisation phase for the FLI had ended.

We measured the effect of this close encounter on each indicator. The three arrows illustrate the difference between the value of the indicator prior to the encounter and its value at the end of the integration. These Deltas show the effect of a close encounter on the final value of the indicators. While the FLI and OFLI react similarly ($\Delta_1 = 5.22$ for the FLI and $\Delta_2 = 4.76$ for the OFLI), the mMEGNO reacts less to this close encounter ($\Delta_3 = 1.65$). It also appears to react slower.
Table \ref{tab:indicateurs} summarises the features of each indicator.

\begin{table*}
    \caption{Specificity of each indicator of chaos. Here `All' signifies that the indicator does not focus on any particular source of chaos.}
    \label{tab:indicateurs}
    \centering
    \begin{tabular}{c c c c}
    \hline \hline
         Indicator & Relativity & Sources of Chaos & Specificity \\
         \hline
         FLI & Relative & All & Very well known \\
         mFLI & Relative & Close encounter & Precise study of one close encounter \\
         OFLI & Relative & All & Correction for differential rotation and quick start \\
         MEGNO & Absolute & All & Oscillations make it ill-suited for a map \\
         mMEGNO & Absolute & All &  Reacts less to close encounters \\ \hline
    \end{tabular}
\end{table*}

All three indicators seem to respond correctly to the evolution of our particles. However, a choice has to be made as to which indicator we use in our analysis. It seems the OFLI is both quicker  than the FLI to reach a first value after the initialisation phase and quicker  than the mMEGNO to react to a close encounter. As we work on short timescales with particles heavily influenced by close encounters, we feel these arguments are sufficient to favour the use of the OFLI. 

\subsubsection{Formula}\label{subsubsec:formula}

The vector $\vec{X}$ represents the state vector (position and velocity) of a particle. We name $f$ the so-called force function that describes the evolution of $\vec{X}$. The tangent vector $\vec{w}$ plays a crucial role in the computation of FLIs. Its evolution is described by:
\begin{equation}
    \begin{cases}
    \dot{\vec{X}} = f(\vec{X}, t) $ with $ \vec{X}(t_0) = \vec{X_0}, \\
        \dot{\vec{w}} = \frac{\partial f}{\partial \vec{X}}(\vec{X},t) . \vec{w} $ with $ \vec{w}(t_0) = \vec{w_0}.
        \end{cases}
\end{equation}

Specifically, the OFLI rests on $\vec{w_2}$, the orthogonal part of $\vec{w}$ with respect to the variational flux:
\begin{equation}
\vec{w_2} = \vec{w} - \frac{(\dot{\vec{X}} . \vec{w}) \dot{\vec{X}}}{||\dot{\vec{X}}||^2}.
\end{equation}

And finally, we have
\begin{equation}
OFLI(t) = \max_{\tau < t}(\ln \Vert \vec{w_2} \Vert).
\end{equation}

The evolution of the vector $\vec{w}$ was computed alongside the evolution of the particles. The initial vector $\vec{w_0}$ was chosen perpendicular to the flux, as \citet{Lega_Froeschle_2001} recommend, and was derived from the gradient $\vec{g}$ of the two-body problem Hamiltonian and the initial state vector $\vec{X_0}$ of the particle studied:
\begin{equation}
    \begin{split}
        \vec{u} & = \vec{g} - \frac{(\dot{\vec{X_0}} . \vec{g}) \dot{\vec{X_0}}}{||\dot{\vec{X_0}}||^2}, \\
        \vec{w_0} & = \frac{\vec{u}}{||\vec{u}||}.
    \end{split}
\end{equation}

\subsection{Computational method}\label{subsec:prog}

The integrator chosen was the RADAU order 15 from \citet{Everhart_1985}. The RADAU is characterised in part by its automated computation of the length of each time increment. We used the ephemeris INPOP from IMCCE \citep{Fienga_al_2009} and added non-gravitational forces \citep[Poynting-Roberston drag and solar radiation pressure; see e.g.][]{Vaubaillon_al_2005}.

The Geminid meteor shower is a well-known shower, dynamically stable compared to streams from Jupiter-family comets. This allows us to test our reasoning. We generated a high ($>10^3$) number of particles with orbits similar to the Geminids. The particles were described by the initial time $t_0$, their state vector $\vec{X_0}$ at $t_0$, and their radius $r$. We assumed a density of $\rho = 1000\text{ kg}/\text{m}^3$ to compute the mass of the particle. We chose the year 2000 A.D  as the initial time, which corresponds to the Geminids current orbit. 

Two sets of initial conditions were processed. In both cases, the mean anomaly was chosen randomly between 0° and 360° in order to evaluate the impact of this parameter on the chaos map. The first set (IC1) was composed of 100080 particles, chosen so as to be relatively close to the ejection conditions of the Geminids (see Table \ref{tab:CI1}). The goal was to simulate a set of initial conditions just large enough to encompass the usual orbits of Geminids. The second set (IC2, Table \ref{tab:CI2}) mapped a larger part of the phase space, and was composed of 99720 particles. Thanks to this second set, we obtained a broader view of the model and investigated what happens on the borders of the Geminid stream. 

For both sets, each heliocentric orbital element of each particle was picked randomly in a chosen interval, as described in Tables \ref{tab:CI1} and \ref{tab:CI2}. This means that no orbital element is fixed: they are all randomly chosen according to a uniform distribution. Such an approach was taken for example by \citet{Todorovic_Novakovic_2015}. Contrary to a uniform cartesian mesh, this method avoids the introduction of a parameter, the step of the mesh. However, as seen in \citet{Gkolias_al_2016}, this non-uniform distribution might blur some details, and so we performed an integration with mesh-like initial conditions. We did not find any improvement in the maps drawn from such initial conditions, and so they are not presented here.

\begin{table}
    \caption{Range of each heliocentric element of IC1 (500 years of integration)}
    \label{tab:CI1}
    \centering
    \begin{tabular}{c c c }
    \hline \hline
         Element & Min & Max \\ 
         \hline
         a (AU) & 1.25 & 1.3 \\
         e & 0.888 & 0.892 \\
         i (°) & 21.675 & 22.675 \\
         $\omega$ (°) & 321.5 & 322.5 \\
         $\Omega$ (°) & 265.03 & 266.03 \\ 
         \hline
    \end{tabular}
\end{table}

\begin{table}
    \caption{Range of each heliocentric element of IC2 (1000 years of integration)}
    \label{tab:CI2}
    \centering
    \begin{tabular}{c c c}
    \hline \hline
         Element & Min & Max \\
         \hline
         a (AU) & 1.2 & 1.35 \\
         e & 0.8 & 0.95 \\
         i (°) & 20 & 24 \\
         $\omega$ (°) & 320 & 323 \\
         $\Omega$ (°) & 264 & 267 \\ 
         \hline
    \end{tabular}
\end{table}

The particles were integrated for 500 years for IC1 and for 1000 years for IC2. We did not need to integrate them further, because we were not simulating the entire lifespan of the Geminids. The evolution of the position, speed, and OFLI of the  particles were recorded, as well as their close encounters with planets (here, mainly the Earth). The encounters were detected when the distance between the particle and the planet was smaller than its Hill radius.

First, we worked with large particles (radius chosen randomly between 10 and 100 mm), on which non-gravitational forces (NGFs) have a negligible effect. We named these data sets IC1 BIN10100 and IC2 BIN10100, depending on the initial conditions used. Then, we also investigated the effect of the NGFs. For this purpose, we replicated IC1 and IC2, changing only the radius. It was picked randomly between 1 and 10 mm (BIN110) and then between 0.1 and 1 mm (BIN011).
We obtained six sets of particles: IC1 BIN10100, IC2 BIN10100, IC1 BIN110, IC2 BIN110, IC1 BIN011, and IC2 BIN011, which are summarised in Table \ref{tab:sets}.

\begin{table}
    \caption{Description of the six different sets of particles. See Tables \ref{tab:CI1} and \ref{tab:CI2} for explanation of IC1 and IC2.}
    \label{tab:sets}
    \centering
    \begin{tabular}{c c c c} 
    \hline \hline
          & Small Particles & Medium Particles & Large Particles \\
          & (0.1 - 1 mm) & (1 - 10 mm) & (10 - 100 mm) \\ 
          \hline
         IC1 & IC1 BIN011 & IC1 BIN110 & IC1 BIN10100 \\
         IC2 & IC2 BIN011 & IC2 BIN110 & IC2 BIN10100 \\ 
         \hline
    \end{tabular}
\end{table}

\section{Results}\label{sec:results}
Maps are usually drawn as a function of initial orbital elements. In our case, maps drawn as a function of the initial semi-major axis and the initial eccentricity ($a$, $e$) of the particles are the only ones presenting distinctive features, and these are therefore the only ones discussed. In these maps, we only plot the value of semi-major axis and eccentricity for each particle, but the value chosen randomly for each of the other elements is not presented.

As explained in the previous section, we tested the whole range of possible values for the mean anomaly. Maps drawn from this element are completely uniform, and so the mean anomaly does not seem to impact the Geminids chaoticity.

\subsection{Resonances}\label{subsec:resonances}

\begin{figure}
    \includegraphics[scale = 0.5]{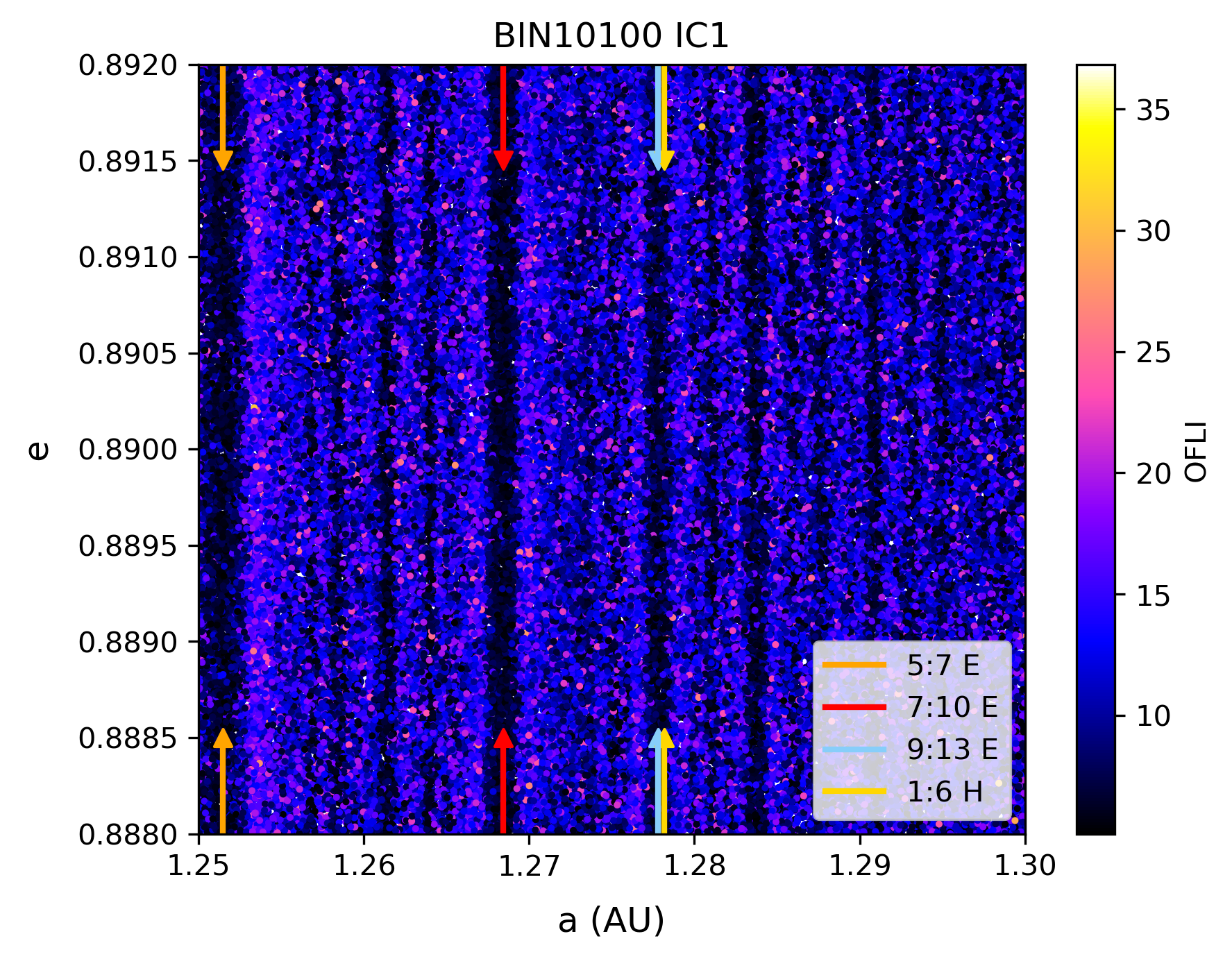}
    \includegraphics[scale = 0.5]{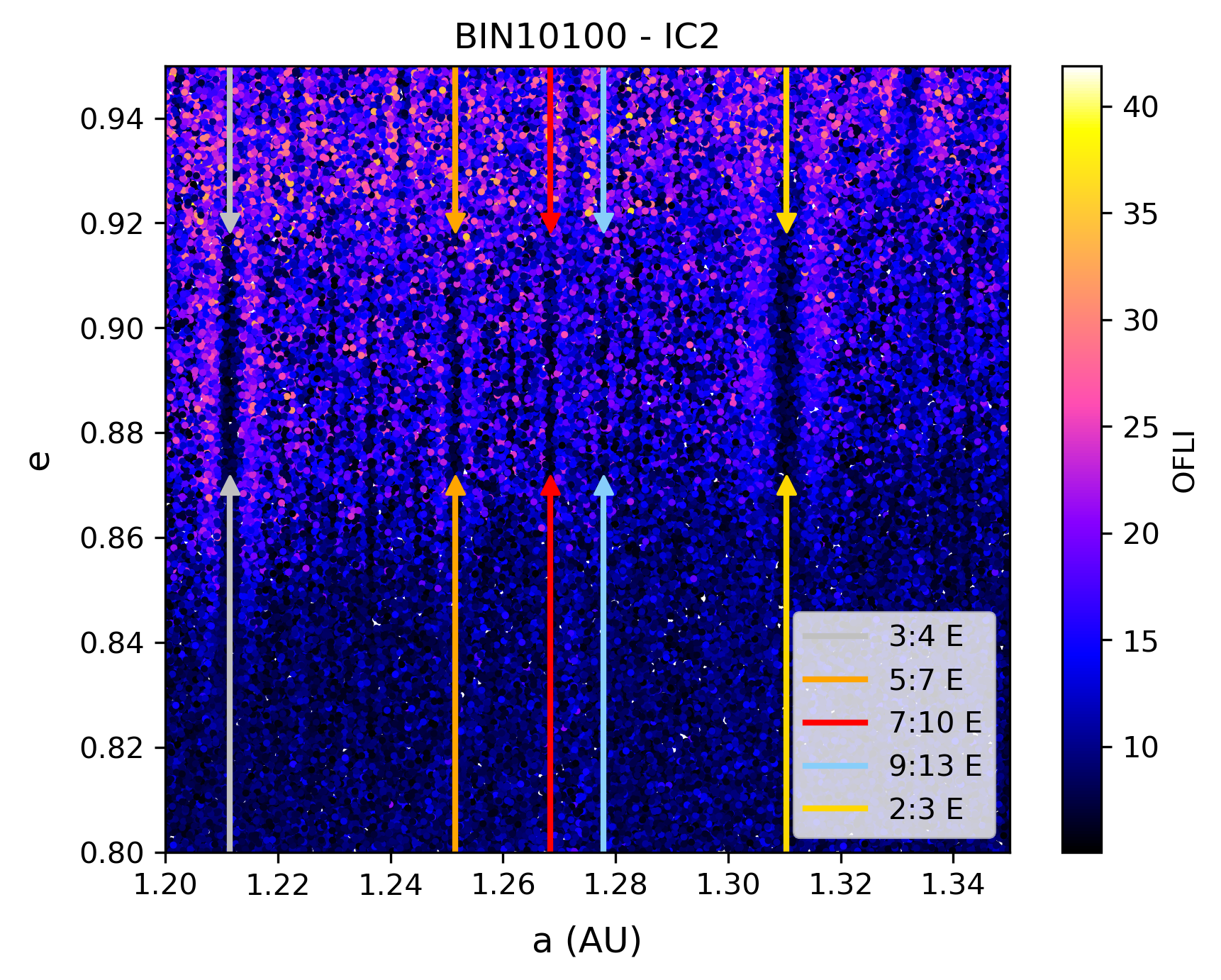}
    \caption{Maps from BIN10100. The arrows point to the dark lines visible in the maps. The colours of the arrows are related to the mean-motion resonances responsible for the line. One map is drawn from the IC1 data set and the other from the IC2 data set (see titles).}
    \label{fig:reso_BIN10100}
\end{figure}

The first maps we drew (Fig. \ref{fig:reso_BIN10100}) are from IC1 BIN10100 and IC2 BIN10100. We note the difference in colour scale between the two data sets: the chaos keeps rising after 500 years. The smallest values are similar, showing the long-term stability of some particles.

The maps present several dark vertical lines, where the chaoticity is much lower. Those lines are a perfect match to the mean-motion resonances (MMRs) listed in Table \ref{tab:reso}. Three of them are mostly present in the BIN10100 IC1 map, and are slightly visible in BIN10100 IC2. Two more (the 2:3 and 3:4 with the Earth) only appear in the BIN10100 IC2 map. 

Two different MMRs fit the dark line at around 1.27 AU: the resonance 1:6 with Mercury and the resonance 9:13 with the Earth. As the Earth plays a determinant role in the evolution of the Geminids, the resonance with the Earth seems more likely. The line is also very thin, which matches with the low order of the resonance with the Earth. 

\begin{table}
    \caption{MMR that fit with the structures observed}
    \label{tab:reso}
    \centering
    \begin{tabular}{c c c} 
    \hline \hline
         Planet & Order & Semi-major axis \\
          & & (AU) \\ 
          \hline
         Earth & 2:3 & 1.31037 \\
         Earth & 3:4 & 1.21141 \\
         Earth & 5:7 & 1.25146 \\
         Earth & 7:10 & 1.26843 \\
         Earth & 9:13 & 1.27781 \\
         Mercury & 1:6 & 1.27817 \\ 
         \hline
    \end{tabular}
\end{table}

\begin{table}
    \caption{MMR found by Ryabova and how they relate to our chaos maps. "Out" means the resonances cannot be found in our chaos maps as their semi-major axis exceeds the bounds of our study. The other resonances ("In") can be seen either only at high eccentricity ("high e") or in the middle of the map ("middle e").}
    \label{tab:reso_Ryabova}
    \centering
    \begin{tabular}{c c c c} 
    \hline \hline
         Planet & Order & Semi-major Axis & Chaos Map\\ 
          & & (AU) & \\ 
          \hline
         Venus & 1:2 & 1.14821 & Out \\
         Venus & 2:5 & 1.33238 & In (high e) \\
         Venus & 3:7 & 1.27249 & In (high e) \\
         Venus & 4:9 & 1.24201 & In (high e) \\
         Earth & 2:3 & 1.31037 & In (middle e) \\
         Earth & 5:7 & 1.25146 & In (middle e) \\
         Jupiter & 7:1 & 1.42130 & Out \\ 
         \hline
    \end{tabular}
\end{table}

\begin{figure}
    \includegraphics[scale = 0.5]{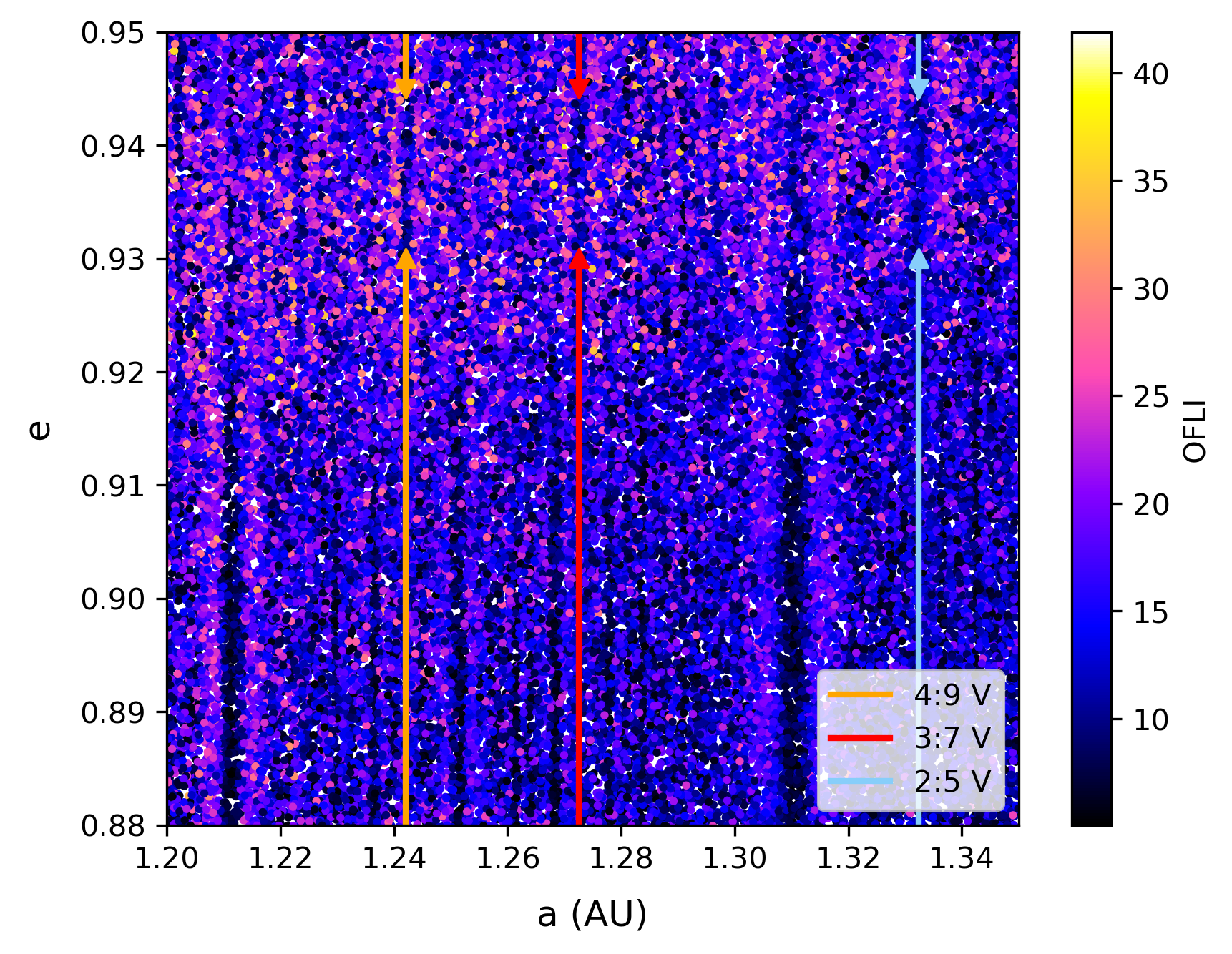}
    \includegraphics[scale = 0.5]{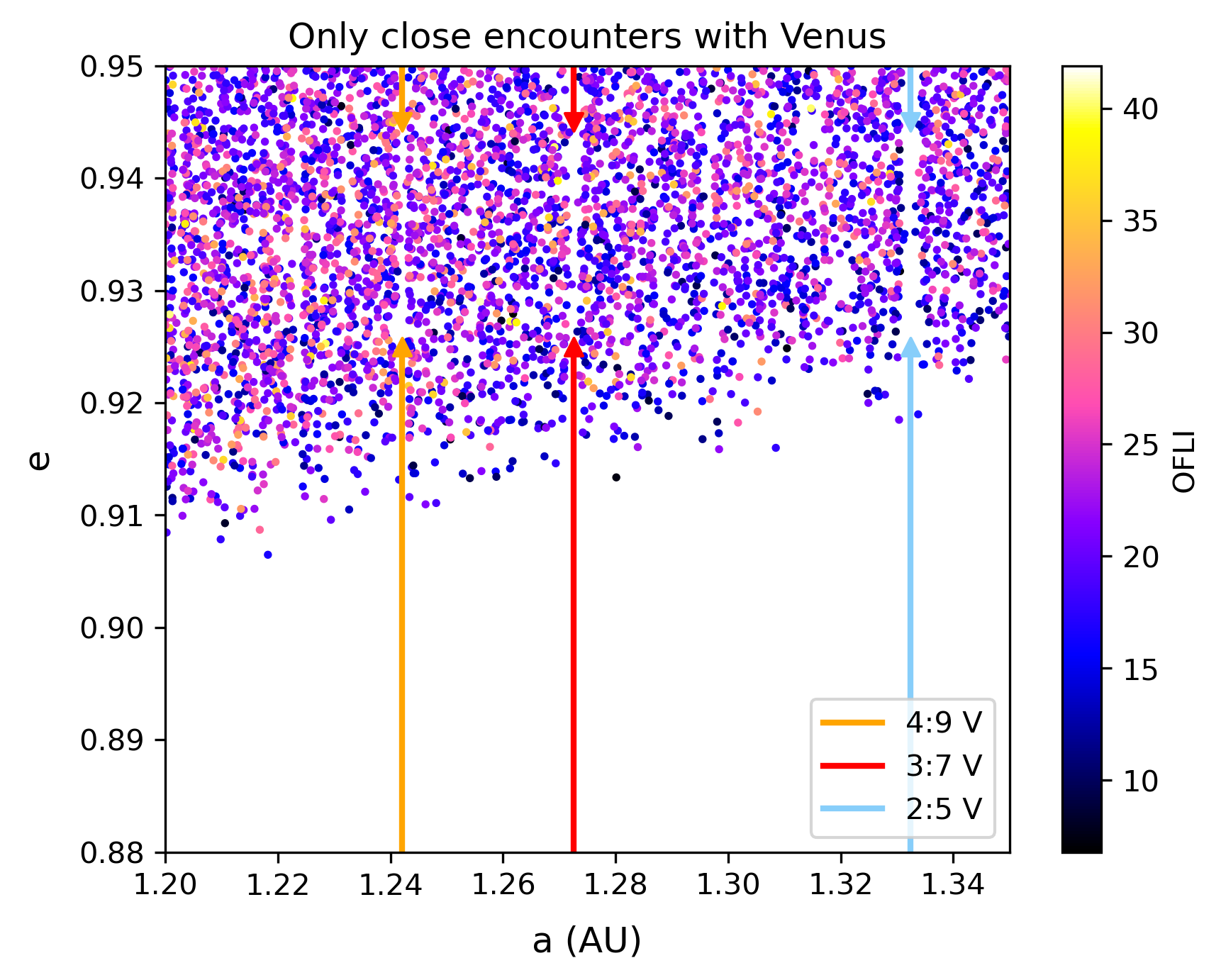}
    \caption{Maps from BIN10100 IC2. On the top panel, we point to some resonances with Venus found by \citet{Ryabova_2022}. On the bottom panel, we only plot particles from BIN10100 IC2 that met with Venus.}
    \label{fig:reso_BIN10100_IC2_venus}
\end{figure}

\citet{Ryabova_2022} studies resonances that play a role in the evolution of the Geminids. The resonances she finds are listed in Table \ref{tab:reso_Ryabova}. As noted in Table \ref{tab:reso_Ryabova}, some of them are outside the bounds of our study, but the 2:3 and 5:7 MMRs with the Earth are detected in the chaos maps.
As for the 2:5, 3:7, and 4:9 MMRs with Venus found by \citet{Ryabova_2022}, they are also visible in our chaos maps, although less clearly: they appear only at high eccentricity, as can be seen in Fig. \ref{fig:reso_BIN10100_IC2_venus}. 

\begin{figure}
    \includegraphics[scale = 0.5]{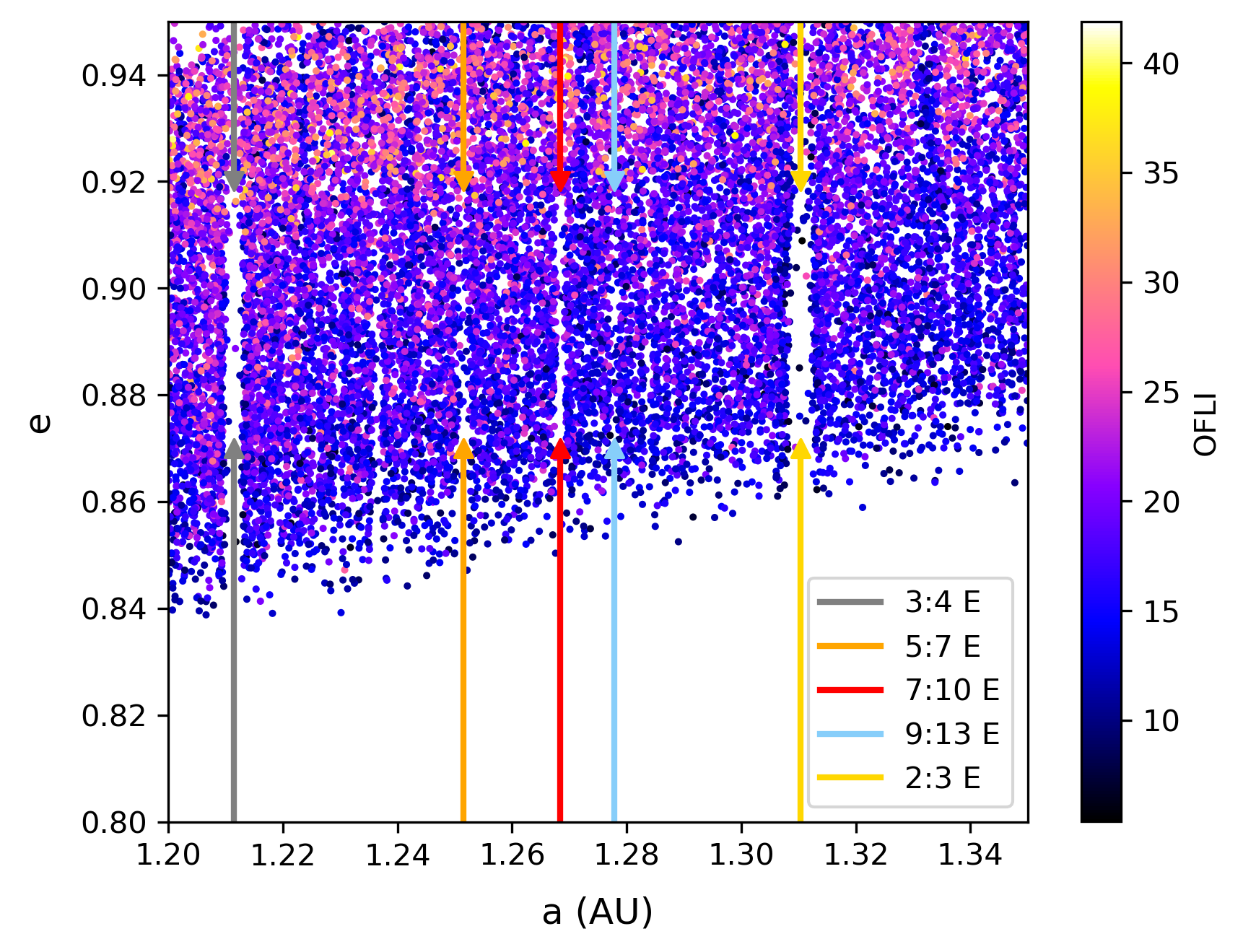}
    \caption{Map from BIN10100 IC2. We only plot particles that met with the Earth.}
    \label{fig:reso_BIN10100_IC2}
\end{figure}

Particles initially inside those MMRs are much less chaotic than others. To understand why, in Fig. \ref{fig:reso_BIN10100_IC2_venus} (respectively Fig. \ref{fig:reso_BIN10100_IC2}), we plot only particles meeting with Venus (respectively with the Earth). These plots reveal the mechanism of stabilisation: the particles initially inside the resonance do not meet with the planet considered. A particle inside a MMR with the Earth, for example, will be trapped there and kept from meeting with the Earth. This will maintain its chaoticity at a relatively low level compared to those that do meet with the Earth. In the same way, a particle trapped in a MMR with Venus cannot meet with this planet.

Using our method and in the specific case where the effect of NGFs is negligible, we are able to replicate the findings of \cite{Ryabova_2022}, but we also find additional information on the effect of MMR. This validates our approach.
In Figs. \ref{fig:reso_BIN10100_IC2_venus} and \ref{fig:reso_BIN10100_IC2}, we can also see the effect of eccentricity. At low eccentricity, the particles cannot meet with the Earth or Venus. At higher eccentricity, they are able to meet with the Earth and their chaoticity rises. At even higher eccentricity, the particles can also meet with Venus, and this is where the OFLI reaches its highest value.
We interpret this as finding the lower bound in eccentricity for the Geminid meteor shower: meteoroids with  too low an initial eccentricity will never meet with the Earth and are therefore not part of the shower. This allows us to check the data we already have on Geminids: particles whose eccentricity is too low are probably contaminants from another dynamical origin.

\subsection{Impact of non-gravitational forces}\label{subsec:NGF}

\begin{figure}
    \includegraphics[scale = 0.5]{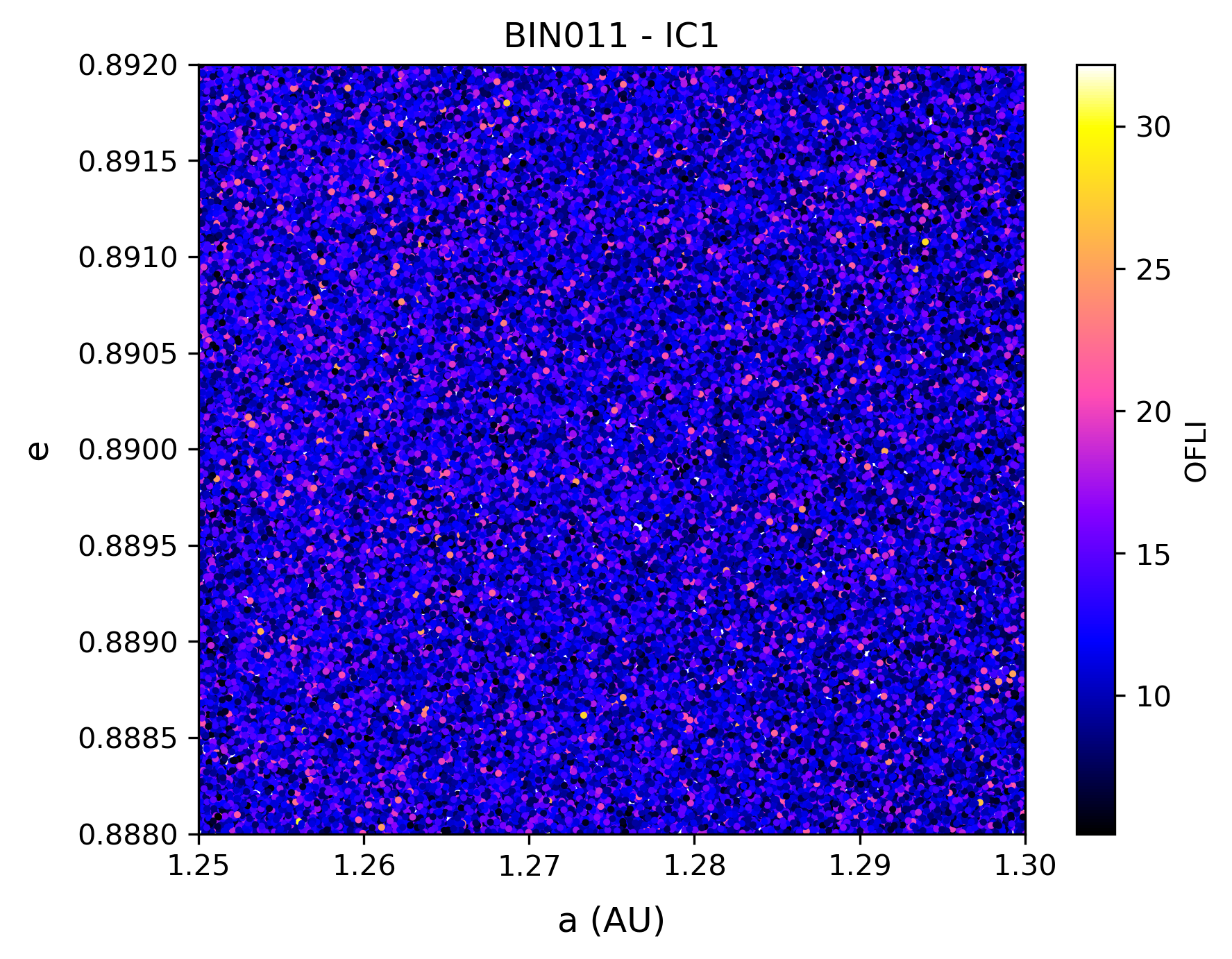}
    \includegraphics[scale = 0.5]{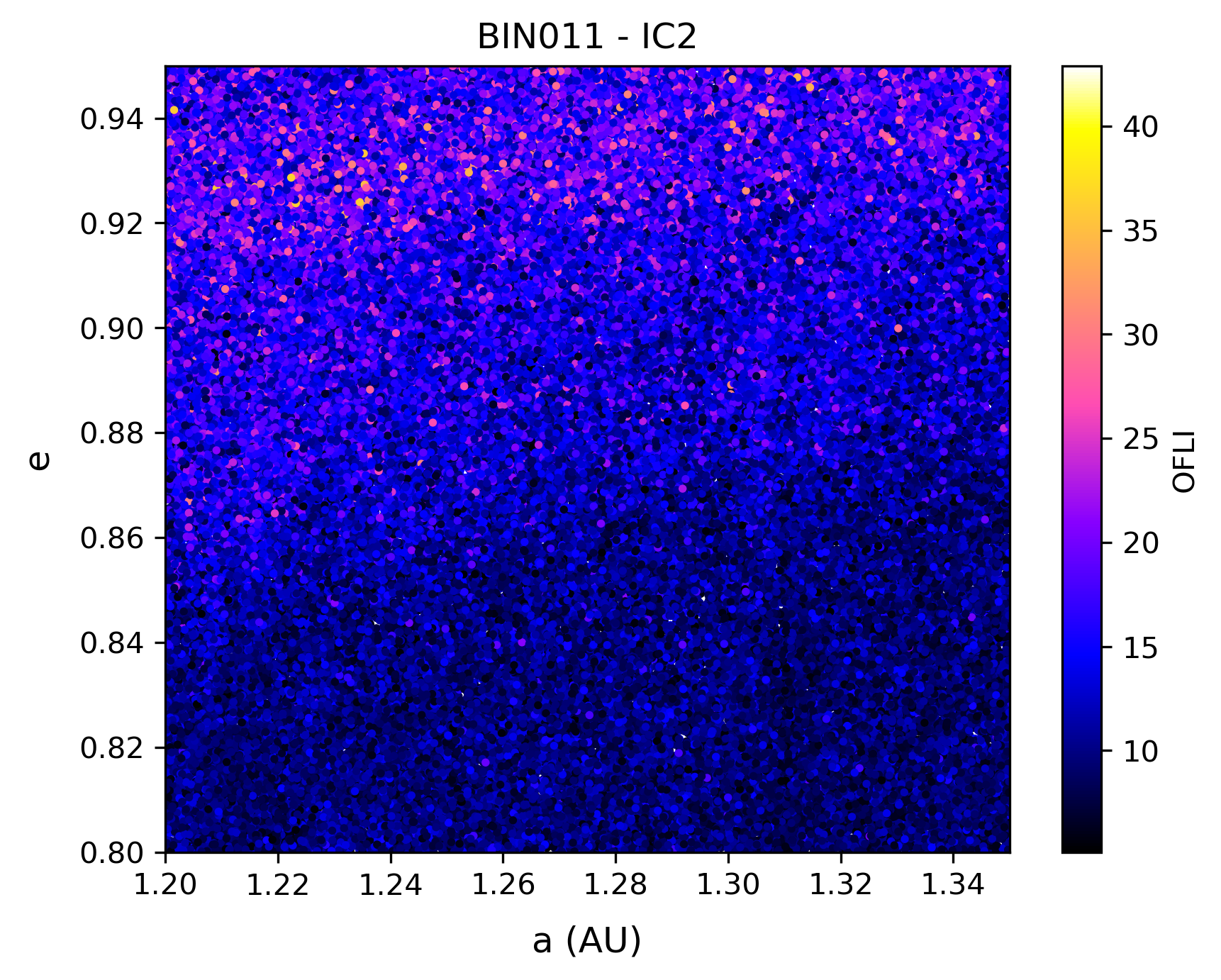}
    \caption{Maps from BIN011 (small particles). One map is drawn from the IC1 data set and one from the IC2 data set (see titles).}
    \label{fig:BIN011}
\end{figure}

To investigate the impact of NGFs, we chose the data sets with smaller radius (BIN110 and BIN011). Maps from BIN110 produce very similar results to the previous section. However, maps from BIN011 (see Fig. \ref{fig:BIN011}) lack the dark vertical lines related to the MMR we analysed; they only present a uniform background for IC1 and the gradient in eccentricity for IC2 (see previous section for explanation). 

Analyses of the interaction between MMRs and NGFs have been conducted before \citep{Liou_Zook_1997}. Here, to better understand this effect, we studied the evolution of a few particles. We chose five particles from BIN10100 IC2 characterised by their initial position with respect to the largest MMR (2:3 and 3:4 with Earth). Particles n°1 and 2 are well outside these MMRs, particle n°3 is close to them but not inside, and particles n°4 and 5 are initially inside the MMRs. We then selected the clones of all of these particles in the BIN110 IC2 and BIN011 IC2 data sets. This allowed us to compare the evolution of particles that differ only from their radius, and thus to measure the influence of MMRs and NGFs for each size. We plotted the evolution of orbital elements and added grey lines that mark the MMRs with the Earth (their respective size corresponds to the size visible in the BIN10100 IC2 map).

\begin{figure}
    \includegraphics[scale = 0.5]{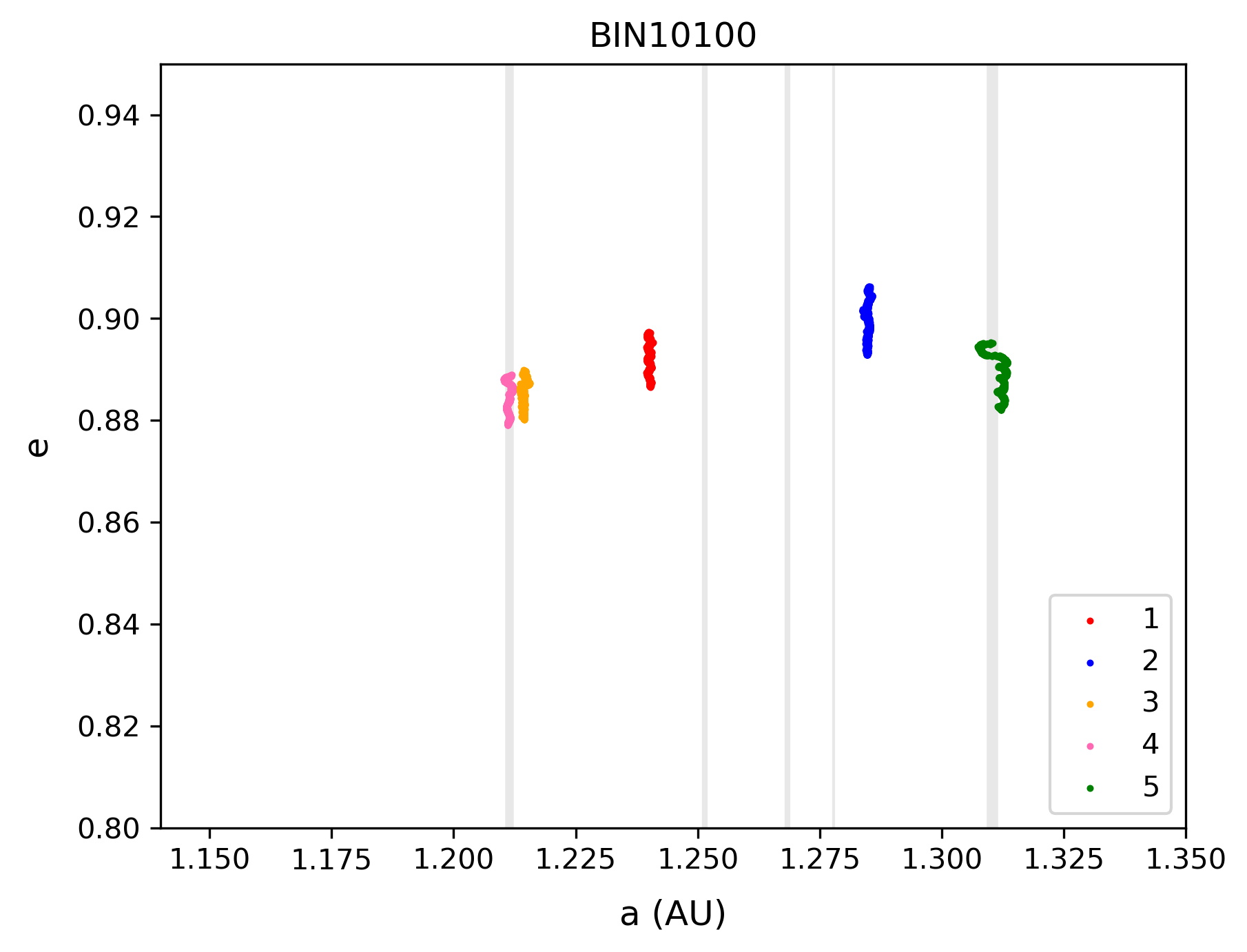}
    \includegraphics[scale = 0.5]{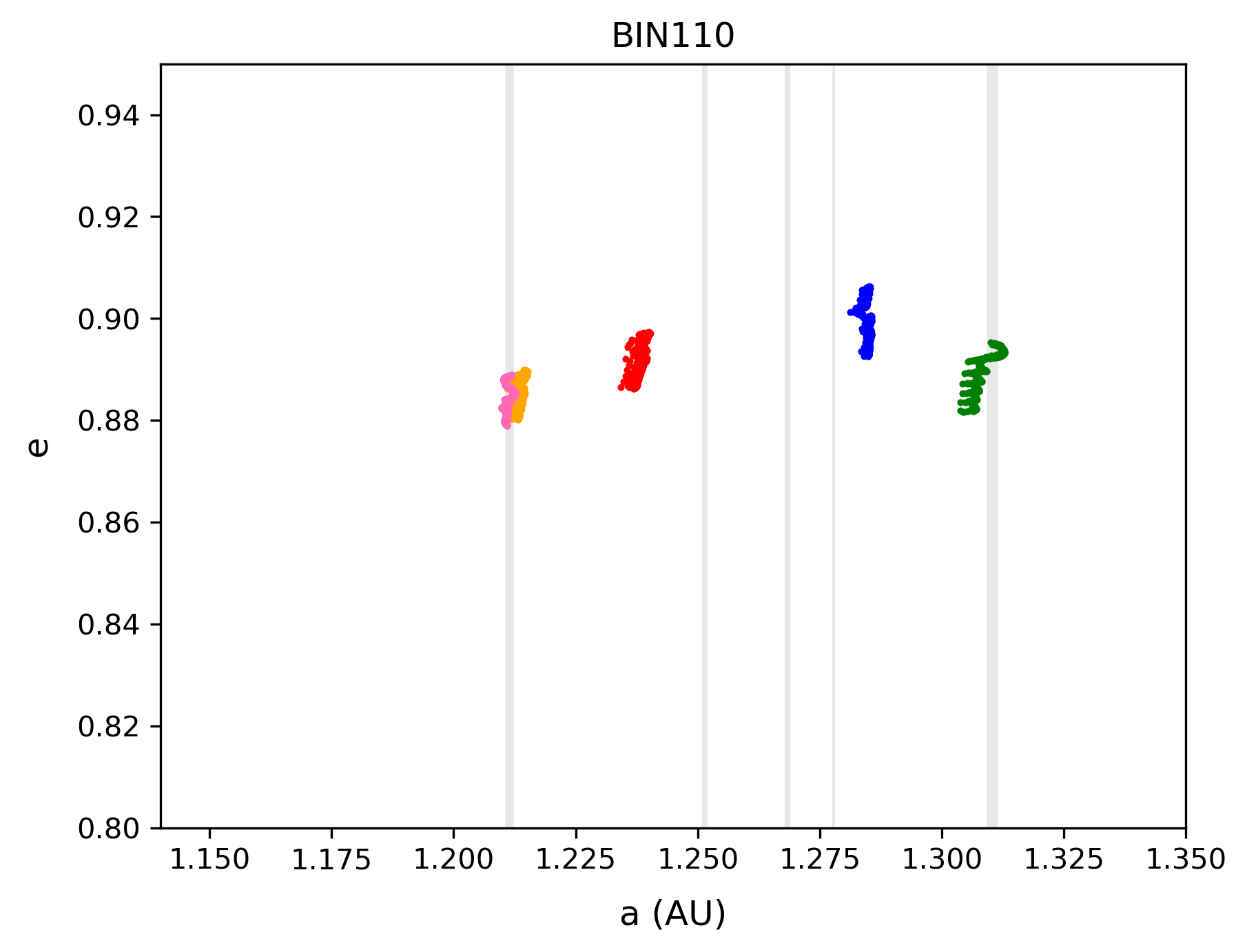}
    \includegraphics[scale = 0.5]{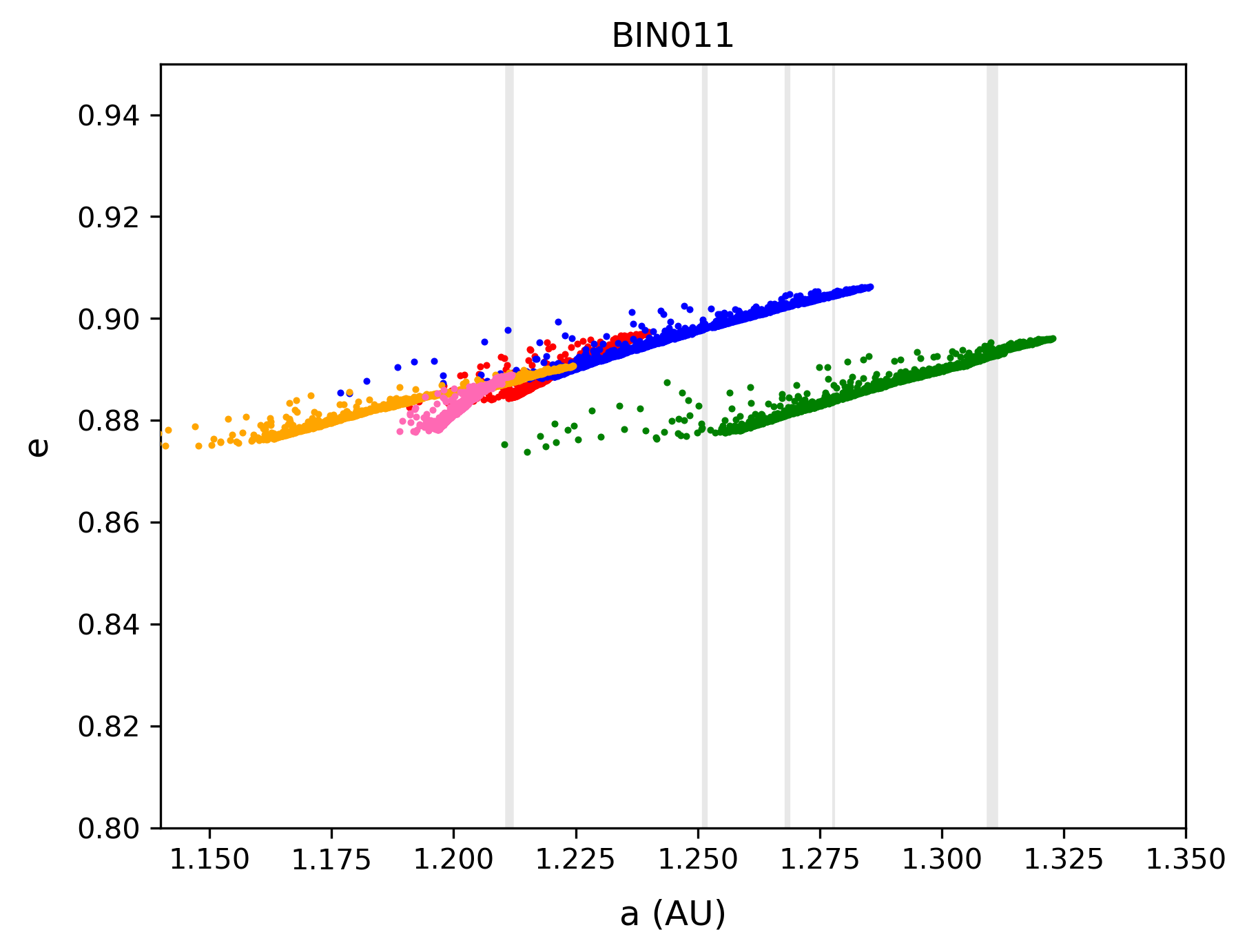}
    \caption{Evolution of five particles, with various radii. We plot the evolution of five particles from BIN10100 IC2 in the first panel and the evolution of their clones from BIN110 IC2 and from BIN011 IC2 in the other two panels, respectively (see titles). The grey lines represent the MMR with the Earth.}
    \label{fig:comp_evol}
\end{figure}

The resulting Fig. \ref{fig:comp_evol} shows the strong diffusion of the small particles from the NGFs, which prevents them from being captured by the MMR. This explains why they do not appear on the BIN011 maps.

Particles of intermediate radius are influenced by MMR but their evolution becomes slightly blurred. This blurry aspect is even more pronounced in the evolution of small particles, with some exotic points that do not seem to be aligned with the general evolution of the orbital elements. This is due to the method of computation of orbital elements: the integration computes the state of each particle as a function of time. The orbital elements are then computed using these data. Doing so, slight changes in the velocity (due to the effect of NGFs) translates into relatively significant changes in semi-major axis $a$. It is well known that $a$ changes drastically when performing such a rough conversion.

In summary, smaller particles lose energy and start to plummet towards the Sun, disregarding any resonances, while larger ones might be locked out of close encounters with the Earth. This has probably a great impact on the distribution of objects in the meteor shower: from Earth, we might see less large objects than originally ejected, because they get captured. The semi-major axes of small objects may also tend to be smaller than those of large objects, even though they originally came from the same parent body.

We wanted to know which value of the radius marks the transition between those two behaviours.
To this end, we used histograms to find the limit radius: the distribution in semi-major axis for the least chaotic particles (final OFLI smaller than 7) present peaks at the MMR semi-major axes for the large particles, while it stays uniform for small particles. We compared the distribution of particles whose radius is smaller than a radius $r_{lim}$ with the distribution of particles whose radius is greater than $r_{lim}$ to check whether or not $r_{lim}$ is indeed the limit radius we are looking for.

\begin{figure}
    \includegraphics[scale = 0.5]{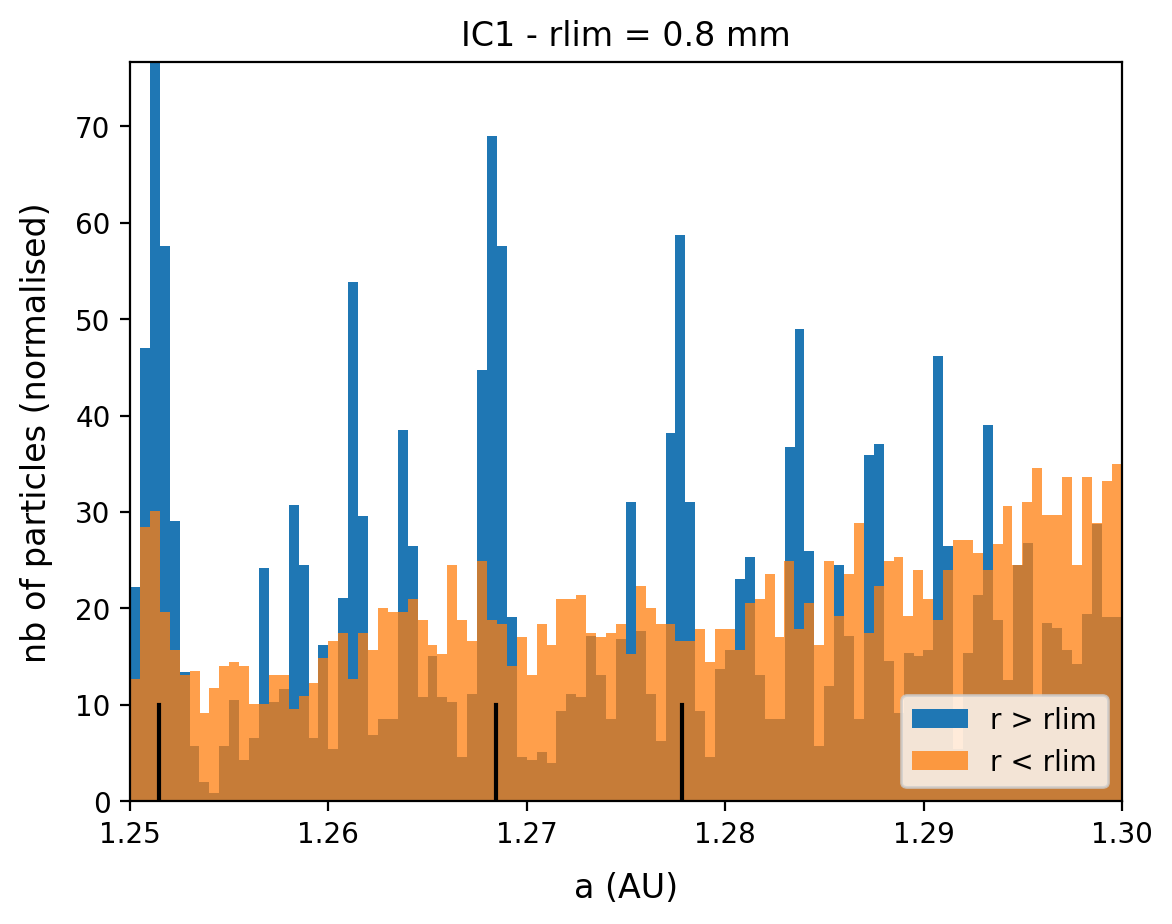}
    \includegraphics[scale = 0.5]{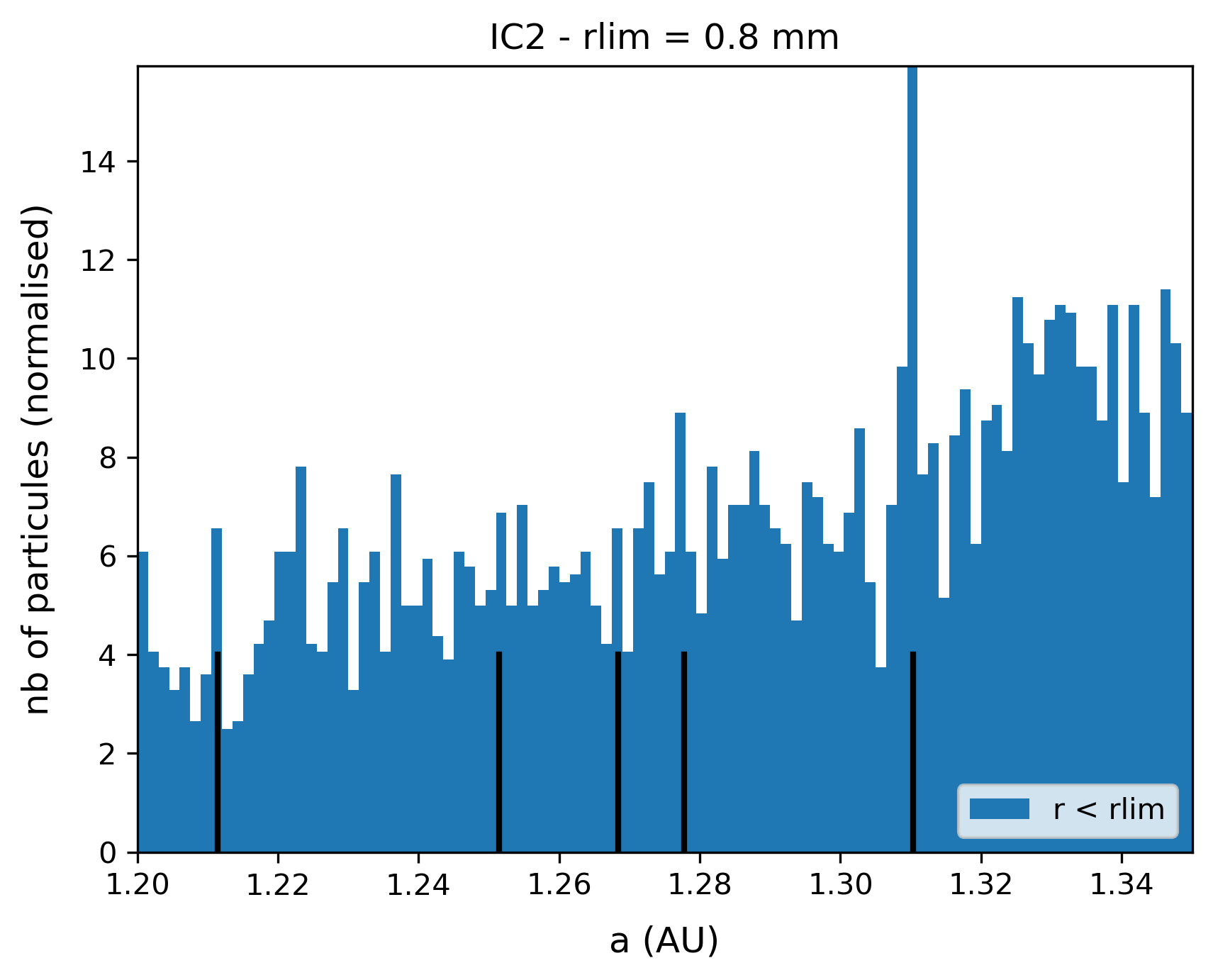}
    \includegraphics[scale = 0.5]{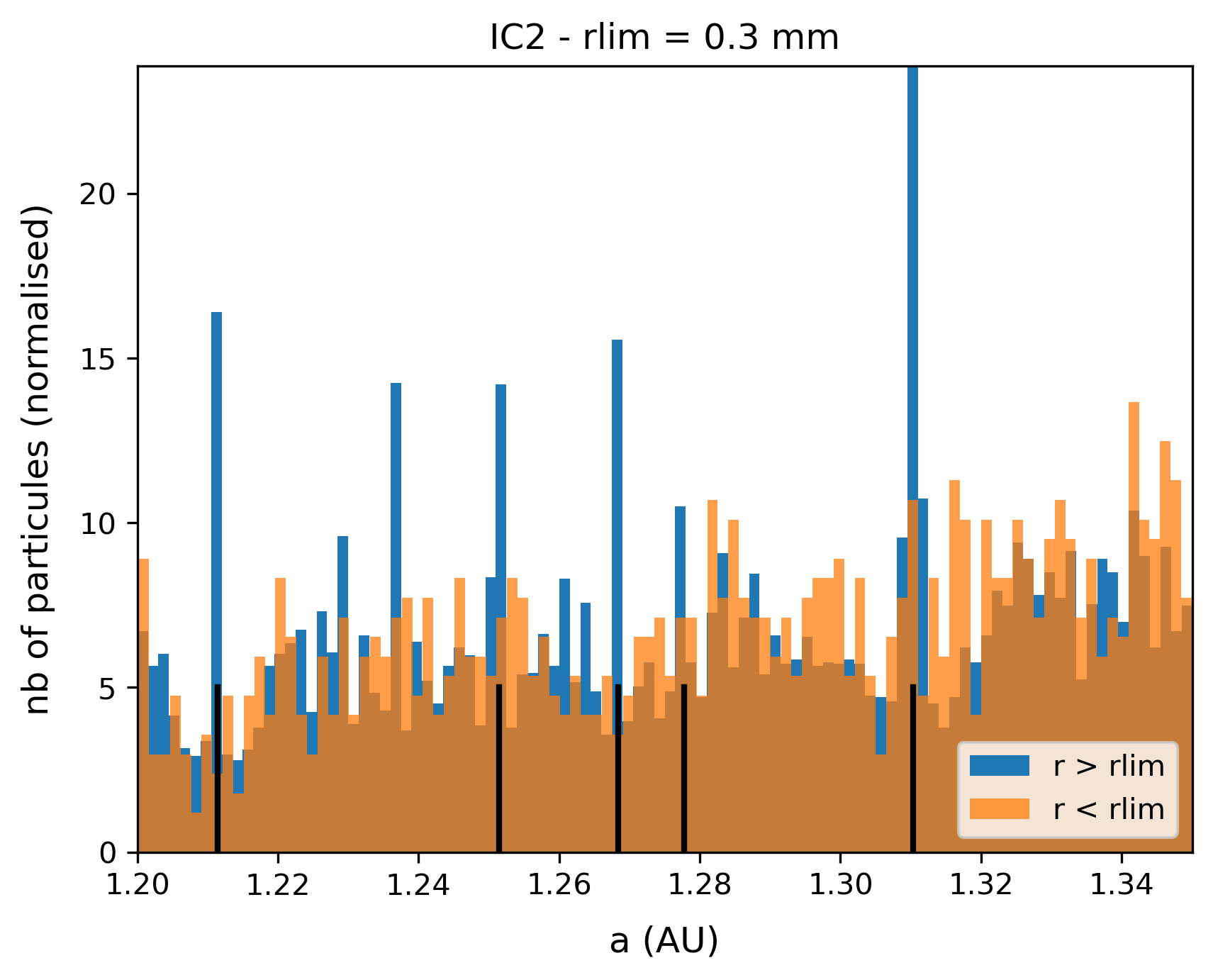}
    \caption{Histograms for the search in radius limit. On each graph, black lines mark the MMRs. The histogram titled `IC1 - $r_{lim} = 0.8$ mm' counts particles from IC1 and compares those whose radius is inferior to $r_{lim}$ with those whose radius is superior to $r_{lim}$. The second histogram counts particles from IC2 with a radius inferior to $r_{lim}$. The last histogram also counts particles from IC2 but changes the value of the radius limit to $r_{lim} = 3.10^{-4}$ m and compares particles whose radius is lower than or superior to $r_{lim}$.}
    \label{fig:lim_radius}
\end{figure}

In Fig. \ref{fig:lim_radius}, with a limit radius of $8.10^{-4}$ m, we obtain the expected result. The second panel in the same figure shows a histogram for IC2 particles, with a radius inferior to the limit chosen. No peak should be visible in this histogram, but one does exist for the strongest MMR in our maps (2:3 with the Earth). This MMR does not appear in the IC1 set, which explains the discrepancy. To make sure the diffusion is stronger than this resonance, $r_{lim}$ must be decreased. On the last panel, we draw a new histogram from IC2 with a limit radius of $3.10^{-4}$ m. This time, the peak on the last MMR disappears for particles whose radius is inferior to this new limit.

The limit radius could therefore be different for other showers, depending on which resonances play a role in their evolution.
We point out that this radius limit is an estimation that could be refined by a detailed theoretical analysis, but such an analysis is beyond the scope of this paper.

\section{Conclusion}

We chose to study chaos in meteoroid streams by drawing chaos maps. Meteoroid dynamics are characterised by a short integration time (of the order of $10^3$ years), many close encounters and potentially strong effect from NGFs, contrary to many previous applications for chaos maps. After analysing some chaos indicators, we find the OFLI to be well suited to our problem. We validated this choice by applying it to the Geminid meteoroid stream. We are able to see the effect of some MMRs on the Geminid meteor shower, obtaining very similar results to those of \citet{Ryabova_2022}. The maps also provided us with interesting insights into how MMRs can trap large particles and prevent them from meeting with planets.

We also show that Geminids are defined by an initial eccentricity higher than approximately $0.84$, because no close encounters with the Earth are found under this value. For even higher values of eccentricity, the particles might meet with Venus in addition to being able to meet with the Earth, adding a new element of chaos.

The effect of NGFs on small meteoroids is also very visible in the chaos maps. We clearly see the effect of diffusion, which completely overpowers the MMR for those small particles. Finally, we computed a first approximation of the radius limit that quantifies the boundary between high diffusion and the effect of MMR. This radius depends on the strength of the MMR and we found $8.10^{-4}$ m and $3.10^{-4}$ m as first approximations. In future works, we may refine them with an analytical study of this phenomenon. 

We also note that the number of large particles in the Geminid meteor shower is probably underestimated given the capture of many of these particles in the MMR. Small particles seem to have a much smaller semi-major axis than large ones when they encounter the Earth, which should be taken into account when looking for parent bodies.

In future works, we will apply our method to other meteor showers, as well as to some meteor groups.

\begin{acknowledgements}
We would like to thank Giovanni Valsecchi, Massimiliano Guzzo and Elena Lega for their interesting insights and advice. Ariane Courtot acknowledges support from the \'{E}cole Doctorale d'Astronomie et d'Astrophysique d'Île-de-France (ED 127).
\end{acknowledgements}

\bibliographystyle{aa} 
\bibliography{biblio.bib}

\begin{thebibliography}{25}
\expandafter\ifx\csname natexlab\endcsname\relax\def\natexlab#1{#1}\fi

\bibitem[{{Barrio}(2005)}]{Barrio_2005}
{Barrio}, R. 2005, Chaos Solitons and Fractals, 25, 711

\bibitem[{{Benettin} {et~al.}(1980){Benettin}, {Galgani}, {Giorgilli}, \&
  {Strelcyn}}]{Benettin_al_1980}
{Benettin}, G., {Galgani}, L., {Giorgilli}, A., \& {Strelcyn}, J.~M. 1980,
  Meccanica, 15, 9

\bibitem[{{Cincotta} {et~al.}(2003){Cincotta}, {Giordano}, \&
  {Sim{\'o}}}]{Cincotta_al_2003}
{Cincotta}, P.~M., {Giordano}, C.~M., \& {Sim{\'o}}, C. 2003, Physica D
  Nonlinear Phenomena, 182, 151

\bibitem[{{Drummond}(1981)}]{Drummond_1981}
{Drummond}, J.~D. 1981, Icarus, 45, 545

\bibitem[{{Egal} {et~al.}(2021){Egal}, {Wiegert}, {Brown}, {Spurn{\'y}},
  {Borovi{\v{c}}ka}, \& {Valsecchi}}]{Egal_al_2021}
{Egal}, A., {Wiegert}, P., {Brown}, P.~G., {et~al.} 2021, \mnras
  [\eprint[arXiv]{2108.00041}]

\bibitem[{{Everhart}(1985)}]{Everhart_1985}
{Everhart}, E. 1985, {An efficient integrator that uses Gauss-Radau spacings},
  ed. A.~{Carusi} \& G.~B. {Valsecchi}, Vol. 115, 185

\bibitem[{{Fienga} {et~al.}(2009){Fienga}, {Laskar}, {Morley}, {Manche},
  {Kuchynka}, {Le Poncin-Lafitte}, {Budnik}, {Gastineau}, \&
  {Somenzi}}]{Fienga_al_2009}
{Fienga}, A., {Laskar}, J., {Morley}, T., {et~al.} 2009, \aap, 507, 1675

\bibitem[{{Fouchard} {et~al.}(2002){Fouchard}, {Lega}, {Froeschl{\'e}}, \&
  {Froeschl{\'e}}}]{Fouchard_al_2002}
{Fouchard}, M., {Lega}, E., {Froeschl{\'e}}, C., \& {Froeschl{\'e}}, C. 2002,
  Celestial Mechanics and Dynamical Astronomy, 83, 205

\bibitem[{{Froeschl{\'e}} {et~al.}(1997){Froeschl{\'e}}, {Lega}, \&
  {Gonczi}}]{Froeschle_al_1997}
{Froeschl{\'e}}, C., {Lega}, E., \& {Gonczi}, R. 1997, Celestial Mechanics and
  Dynamical Astronomy, 67, 41

\bibitem[{{Gkolias} {et~al.}(2016){Gkolias}, {Daquin}, {Gachet}, \&
  {Rosengren}}]{Gkolias_al_2016}
{Gkolias}, I., {Daquin}, J., {Gachet}, F., \& {Rosengren}, A.~J. 2016, \aj,
  152, 119

\bibitem[{{Guennoun} {et~al.}(2019){Guennoun}, {Vaubaillon}, { {\v{C} }apek},
  {Koten}, \& {Benkhaldoun}}]{Guennoun_al_2019}
{Guennoun}, M., {Vaubaillon}, J., { {\v{C} }apek}, D., {Koten}, P., \&
  {Benkhaldoun}, Z. 2019, Astronomy \& Astrophysics, 622, A84

\bibitem[{{Guzzo} \& {Lega}(2015)}]{Guzzo_Lega_2015}
{Guzzo}, M. \& {Lega}, E. 2015, Astronomy \& Astrophysics, 579, A79

\bibitem[{{Jenniskens}(2008)}]{Jenniskens_2008}
{Jenniskens}, P. 2008, Icarus, 194, 13

\bibitem[{{Jopek} {et~al.}(2008){Jopek}, {Rudawska}, \&
  {Bartczak}}]{Jopek_al_2008}
{Jopek}, T.~J., {Rudawska}, R., \& {Bartczak}, P. 2008, Earth Moon and Planets,
  102, 73

\bibitem[{{Laskar}(1990)}]{Laskar_1990}
{Laskar}, J. 1990, Icarus, 88, 266

\bibitem[{{Lega} \& {Froeschl{\'e}}(2001)}]{Lega_Froeschle_2001}
{Lega}, E. \& {Froeschl{\'e}}, C. 2001, Celestial Mechanics and Dynamical
  Astronomy, 81, 129

\bibitem[{{Liou} \& {Zook}(1997)}]{Liou_Zook_1997}
{Liou}, J.-C. \& {Zook}, H.~A. 1997, Icarus, 128, 354

\bibitem[{{Markus}(1990)}]{Markus_1990}
{Markus}, M. 1990, Computers in Physics, 4, 481

\bibitem[{{Rudawska} \& {Jopek}(2010)}]{Rudawska_Jopek_2010}
{Rudawska}, R. \& {Jopek}, T.~J. 2010, in Icy Bodies of the Solar System, ed.
  J.~A. {Fernandez}, D.~{Lazzaro}, D.~{Prialnik}, \& R.~{Schulz}, Vol. 263,
  253--256

\bibitem[{{Rudawska} {et~al.}(2015){Rudawska}, {Matlovi{\v{c}}}, {T{\'o}th}, \&
  {Korno{\v{s}}}}]{Rudaska_al_2015}
{Rudawska}, R., {Matlovi{\v{c}}}, P., {T{\'o}th}, J., \& {Korno{\v{s}}}, L.
  2015, Planetary and Space Science, 118, 38

\bibitem[{{Ryabova}(2022)}]{Ryabova_2022}
{Ryabova}, G.~O. 2022, Planetary and Space Science, 210, 105378

\bibitem[{{Southworth} \& {Hawkins}(1963)}]{Southworth_Hawkins_1963}
{Southworth}, R.~B. \& {Hawkins}, G.~S. 1963, Smithsonian Contributions to
  Astrophysics, 7, 261

\bibitem[{{Todorovi{\'c}} \& {Novakovi{\'c}}(2015)}]{Todorovic_Novakovic_2015}
{Todorovi{\'c}}, N. \& {Novakovi{\'c}}, B. 2015, \mnras, 451, 1637

\bibitem[{{Valsecchi} {et~al.}(1999){Valsecchi}, {Jopek}, \&
  {Froeschle}}]{Valsecchi_al_1999}
{Valsecchi}, G.~B., {Jopek}, T.~J., \& {Froeschle}, C. 1999, \mnras, 304, 743

\bibitem[{{Vaubaillon} {et~al.}(2005){Vaubaillon}, {Colas}, \&
  {Jorda}}]{Vaubaillon_al_2005}
{Vaubaillon}, J., {Colas}, F., \& {Jorda}, L. 2005, Astronomy and Astrophysics,
  439, 751

\end{thebibliography}

\end{document}